\documentclass[american,aps,pre,showpacs,twocolumn,showkeys,10pt]{revtex4-2}
\usepackage{etoolbox}
\usepackage{color}
\makeatletter
\usepackage{babel}
\makeatother

\usepackage{graphicx}%
\usepackage{multirow}%
\usepackage{amsmath,amssymb,amsfonts}%
\usepackage{amsthm}%
\usepackage{mathrsfs}%
\usepackage[title]{appendix}%
\usepackage{xcolor}%
\usepackage{textcomp}%
\usepackage{manyfoot}%
\usepackage{booktabs}%
\usepackage{algorithm}%
\usepackage{algorithmicx}%
\usepackage{algpseudocode}%
\usepackage{listings}%

\newcommand{\avg}[1]{\langle{#1}\rangle}

\makeatletter
\providecommand{\@reinserts}{}
\makeatother

\begin{document}

\title{Extracting the geometric backbone of bipartite networks}

\author{Luc\'ia S. Ram\'irez  $^{1,2}$}
\author{Roya Aliakbarisani$^{1,2}$}
\author{M. \'Angeles Serrano$^{1,2,3}$}
\author{Mari\'an Bogu\~n\'a$^{1,2}$}
\email {marian.boguna@ub.edu}

\affiliation{$1.$ Departament de F\'{i}sica de la Mat\`{e}ria Condensada, Universitat de Barcelona, Mart\'{i} i Franqu\`es 1, 08028 Barcelona, Spain}
\affiliation{$2.$ Universitat de Barcelona Institute of Complex Systems (UBICS), Universitat de Barcelona, Barcelona, Spain}
\affiliation{$3.$ Instituci\'{o} Catalana de Recerca i Estudis Avan\c{c}ats ICREA, Passeig Llu\'{i}s Companys 23, 08010 Barcelona, Spain}


\begin{abstract}
Real bipartite networks combine degree-constrained random mixing with structured, locality-like rules. We introduce a statistical filter that benchmarks node-level bipartite clustering against degree-preserving randomizations to classify nodes as geometric (signal) or random-like (noise). In synthetic mixtures with known ground truth, the filter achieves high F-scores and sharpens inference of latent geometric parameters. Applied to four empirical systems --metabolism, online group membership, plant-pollinator interactions, and languages-- it isolates recurrent neighborhoods while removing ubiquitous or weakly co-occurring entities. Filtering exposes a compact geometric backbone that disproportionately sustains connectivity under percolation and preserves downstream classifier accuracy in node-feature tasks, offering a simple, scalable way to disentangle structure from noise in bipartite networks.
\end{abstract}

\maketitle

\section{Introduction}

Mathematics provides the language of nature, and network science provides the language of complex systems \citep{Newman2010Networks,Barabasi2016NetworkScience}. In such networks, the arrangement of nodes and edges --the topology-- offers a window into how the system forms and operates \citep{Newman2010Networks}. Yet, in most real-world cases, different connectivity mechanisms overlap within the same network. A portion of connections might emerge through a random process (while respecting certain local properties, such as preserving each node's degree) \citep{Newman2001PRE,ParkNewman2004PRE,MaslovSneppen2002Science}, whereas others follow structured principles, such as similarity, hierarchy, or spatial proximity \citep{SB1,Papadopoulos2012Nature,Krioukov2010PRE,Latapy2008SN}. These varied mechanisms, often coexisting, reflect the multiple forces at play in shaping a network's overall structure \citep{KarrerNewman2011PRE,Larremore2014PRE}.

A food web highlights this coexistence of random-like and rule-based connectivity. Some species, known as generalists, can feed on a wide variety of prey, suggesting connections that appear relatively random \citep{WilliamsMartinez2000Nature,May1972Nature,AllesinaTang2012Nature}. Other species, the specialists, focus on narrow sets of closely related prey, following a specific ecological logic that resembles a more deterministic or rule-based mechanism \citep{WilliamsMartinez2000Nature,Krause2003Nature,ThebaultFontaine2010Science}. Similarly, in technological systems --like user-item recommendation networks-- certain items might be recommended broadly regardless of the user profile (akin to a random selection with some constraints), while others are recommended only to users matching specific criteria, reflecting a more structured approach \citep{GomezUribeHunt2015TMIS,Ziegler2005WWW,KaminskasBridge2016TIIS}.

Disentangling these connectivity types is crucial to understanding how the system's structure arises and how it might evolve. ``Random'' does not always mean mere noise: in ecological scenarios, random-like connections can bolster stability or buffer change depending on architecture \citep{May1972Nature,ThebaultFontaine2010Science,AllesinaTang2012Nature}, while in machine learning, random edges might obscure meaningful relationships \citep{Ziegler2005WWW,KaminskasBridge2016TIIS}. For instance, in a recommendation system, random recommendations could help surface novel items but might also reduce relevance if not managed properly \citep{GomezUribeHunt2015TMIS,KaminskasBridge2016TIIS}. Recognizing when random connectivity confers adaptability and when it constitutes unhelpful noise is essential to refining both our conceptual understanding and practical applications.

In this paper, we introduce a methodology for separating random and nonrandom connectivity in bipartite networks. 
Bipartite structures emerge in many fields \citep{Newman2010Networks,Latapy2008SN}: plant-pollinator networks, where different pollinators visit different plants \citep{BascompteJordano2007}; metabolic interactions, linking reactions to metabolites \citep{FellWagner2000NatBiotech,Jeong2000Nature}; scientific collaborations, pairing researchers with their publications \citep{Newman2001PNAS}; user-item recommendation systems, matching people with products or services \citep{KorenBellVolinsky2009,Ricci2011Handbook}; and node-feature graphs in machine learning, relating data instances to their attributes \citep{Dhillon2001KDD}. By isolating and quantifying the random component in these systems, our method offers a clearer view of the underlying forces that shape network organization. This, in turn, facilitates more accurate modeling, enabling predictions of how the network might respond to changes --such as the introduction or removal of nodes-- and aiding in the design of interventions or optimizations that take both random and structured processes into account \citep{Tumminello2011PONE,Saracco2015SciRep,Saracco2017NJP,Neal2013SNAM}.

Ultimately, understanding the balance between random and nonrandom connectivity can deepen our appreciation of how complex systems emerge, adapt, and function. By identifying and disentangling these dual mechanisms, we become better equipped to analyze, predict, and influence the behavior of a wide range of bipartite networks, from ecological webs to machine-learning pipelines \citep{KarrerNewman2011PRE,Larremore2014PRE}.

\begin{figure*}
\centering    
{\includegraphics[width=2\columnwidth]{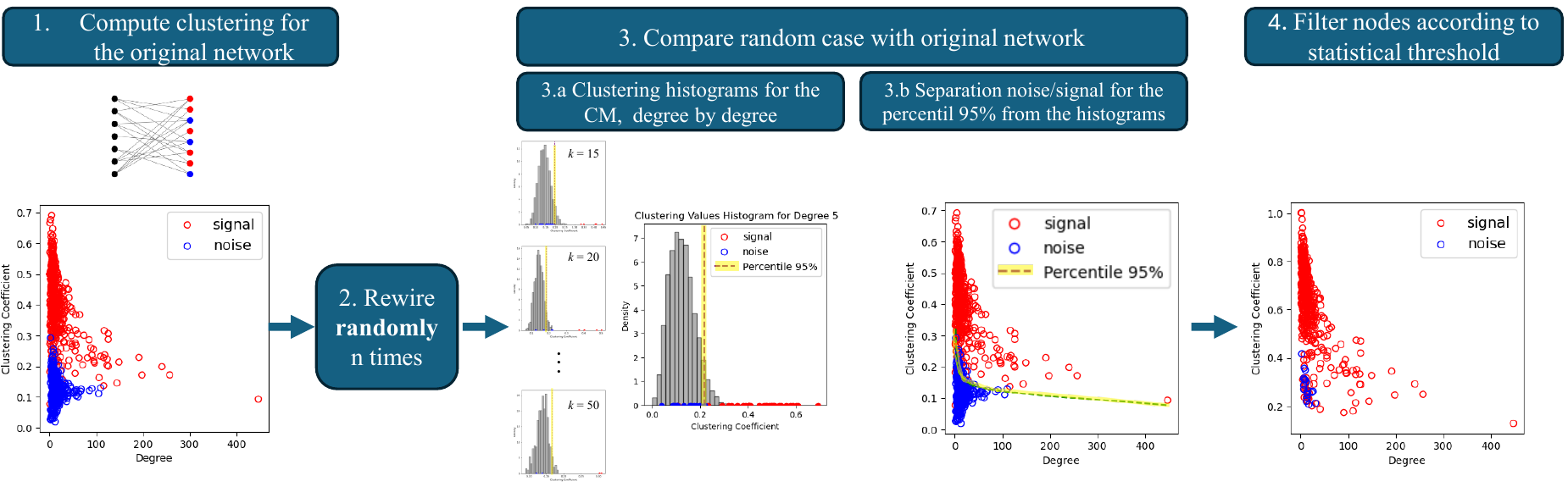}}
\caption{
\textbf{Pipeline for filtering noisy nodes in bipartite networks.} The process involves four main steps: (1) computing the clustering coefficient for the original network, (2) rewiring the network randomly $n$ times, (3) generating clustering histograms for the configuration model (CM), degree by degree, and comparing them to the clustering of the original network using a statistical threshold (e.g., significance level $p= 0.05$, or percentile $95\%$), and (4) filtering nodes based on the comparison. Nodes with clustering coefficients above the yellow line in step $3$ are considered signal, while those below the threshold are classified as noise. 
}
\label{fig:pipeline}
\end{figure*}

\section{Results}

As discussed above, in many real-world bipartite networks, edges emerge through multiple mechanisms that coexist and overlap \citep{Newman2001PRE,MolloyReed1995RSA,MolloyReed1998CPC}. Specifically, here we consider a network composed of two distinct sets of nodes: type A and type B. For some nodes of type B, their connections to nodes of type A appear to follow a random process, one that preserves each node's individual degree but otherwise distributes edges as freely as possible --the so-called configuration model (CM) \citep{MaslovSneppen2002Science,RobertsCoolen2012PRE,Strona2014NatComm,Saracco2015SciRep,Saracco2017NJP}. This random mechanism explains a portion of the network in which no clear pattern or rule-based structure dictates connections, except that the total number of connections per node is maintained. However, such random connections alone do not fully capture the network's structural complexity.

The rest of the type B nodes exhibit a non-random linkage pattern. That is, their connections to nodes in type A follow principles or constraints beyond the configuration model, producing topological features and patterns that deviate from a purely random arrangement \citep{Tumminello2011PONE,Milo2002PNAS,Kitsak2017PRE}. Although we do not specify here what exact processes drive this non-random behavior, we do know it must be distinguished from the degree-constrained random baseline if we are to understand the system comprehensively. The challenge, therefore, is to disentangle these two connectivity modes --random and non-random-- within the same bipartite network. By doing so, we can determine how much of the overall structure is attributable to a random mechanism and identify the specific structural signatures of the non-random connections, illuminating the multiple forces that shape network organization.

To achieve this goal, we first select a topological property capable of distinguishing random from non-random connectivity. The degree of individual nodes alone cannot serve this purpose, as degree is a local property that can be satisfied even in highly random cases. Instead, we use the bipartite clustering coefficient, which reflects the density of ``squares'' formed by two type A nodes and two type B nodes (see {\it Materials and Methods}) \citep{Latapy2008SN,Zhang2008PhysicaA,Opsahl2013SN}. Clustering is crucial here because it often indicates underlying similarity or dissimilarity, suggesting the presence of a latent similarity space shaping the network's topology \citep{Papadopoulos2012Nature,Krioukov2010PRE,Kitsak2017PRE}. Importantly, the clustering coefficient can also be measured at the level of individual nodes, allowing the identification of nodes with particularly high or low clustering values.

However, a single node's clustering coefficient is strongly influenced by its degree, making such a selection difficult when the network has heterogeneous degree distributions. To address this, we compare clustering within degree classes \citep{SofferVazquez2005PRE}. Specifically, we generate randomized versions of the network by rewiring edges while preserving the overall degree sequence \citep{MaslovSneppen2002Science,RobertsCoolen2012PRE,Strona2014NatComm}. Then, for each degree class, we compute the distribution of clustering coefficients across a large number of randomized networks and assess each node's observed clustering relative to this random baseline \citep{Tumminello2011PONE}. If a type B node's clustering coefficient significantly exceeds the random expectation for its degree class, we classify its connections as non-random; otherwise, the node is classified as compatible with randomness.

\subsection{Node Filtration} 
\label{sec:filter}

Figure~\ref{fig:pipeline} sketches the pipeline of our proposal to filter out noisy nodes of a real bipartite network. It is divided in three main components that we describe below.
\begin{itemize}
\item
{\bf Step 1. Compute node-level bipartite clustering in the original network.}
Multiple clustering definitions exist for bipartite graphs; here we adopt the measure in {\it Materials and Methods} (\eqref{eq:clustering}), which is well suited to distinguishing geometric (signal) from random-like (noise) structure. We compute $c_{i}$ for all nodes in set $B$ and plot clustering versus degree (Step~1 in Fig.~\ref{fig:pipeline}). In the illustrative synthetic example --a mixture of the Configuration Model and the geometric $\mathbb{S}^1$ model~\cite{SB1,SB2}-- points labeled as noise (blue) lie systematically below geometric points (red).
\item
{\bf Step 2. Rewiring the network}. We compare the observed clustering with degree-preserving nulls built by random edge swaps. At each step, choose two $A$--$B$ edges $(i,j)$ and $(k,\ell)$ and, if no multiedge would form, rewire to $(i,\ell)$ and $(k,j)$; this preserves both degree sequences and realizes the configuration model. We perform enough swaps to rewire every edge at least once and generate $1000$ CM surrogates. For each surrogate we measure the clustering of type-$B$ nodes and, across surrogates, compile the degree-conditioned empirical null distributions (histograms) used in Step~3a of Fig.~\ref{fig:pipeline}.

\item
\textbf{Step 3. Comparing the CM with the original network.} Using the degree-conditioned histograms from the CM surrogates as the null, we set an upper-tail clustering threshold for each degree class at a chosen significance level $p$ (e.g., $p=0.05$, the 95th percentile; Fig.~\ref{fig:pipeline}). Nodes whose observed clustering exceeds this threshold are labeled geometric (signal); those below are compatible with the CM and labeled noise. The CM serves only as the random-mixing baseline---we make no assumptions about the signal beyond exhibiting clustering in excess of the null. The resulting per-degree thresholds trace a separating curve in the clustering--degree plane (Step~3.b in Fig.~\ref{fig:pipeline}).
\end{itemize}

\subsection{Experiments in synthetic bipartite networks}

\begin{figure}[t]
\centering    
{\includegraphics[width=\columnwidth]{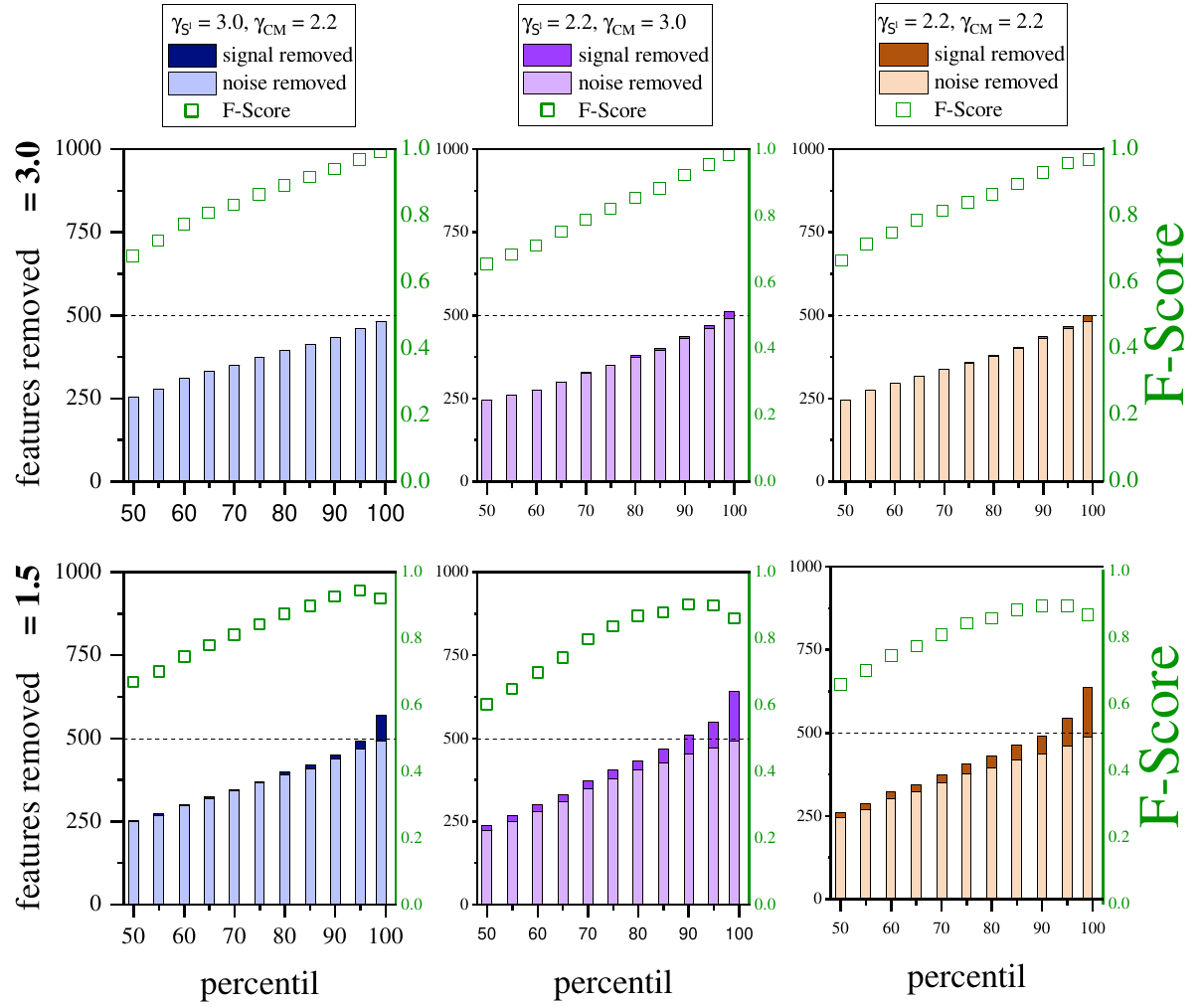}}
\caption{
\textbf{Performance of noise detection in synthetic bipartite networks.} The panels report the number of nodes classified as noise across different percentiles --corresponding to different significance levels $p$. Light bars denote true positives --noisy nodes correctly identified--whereas dark bars denote false positives --signal nodes misclassified as noise. Each row corresponds to a different value of $\beta$ (indicated in the figure), and each column represents combinations of $\gamma_{\mathbb{S}^1}$ and $\gamma_{CM}$ for the degree distributions of sets $B^{\mathbb{S}^1}$ and $B^{\mbox{\tiny{CM}}}$. Green squares show the F-Score at each percentile, capturing the precision-recall trade-off. For percentiles above $0.85\%$, F-Score values are typically $\gtrsim 0.9$, demonstrating the method's robustness.
 }\label{fig:results_synthetic}
\end{figure}

We first test our method in a fully controlled environment where we have access to the network's ground truth. We then define an ensemble of synthetic bipartite graphs by combining two well-known random graph ensembles: the soft version of the bipartite configuration model (SCM) and the bipartite-$\mathbb{S}^1$ model. The SCM is the ``canonical'' version of the CM, where the degree sequence is fixed on average ---each node is assigned an expected degree $\kappa$ rather than its actual degree. Hence, the SCM is fully characterized by the hidden degree distributions for both node types, $\rho_{\mbox{\tiny{A}}}(\kappa_{\mbox{\tiny{A}}})$ and $\rho_{\mbox{\tiny{B}}}(\kappa_{\mbox{\tiny{B}}})$.

The bipartite-$\mathbb{S}^1$ model~\cite{Serrano:2012nh,Kitsak2017PRE} is the bipartite counterpart of the $\mathbb{S}^1$ model~\cite{SB1,SB2}. In this model, nodes are assigned expected degrees according to given distributions and lie on a metric space (a $D$-sphere with $D=1$ in this case). The probability of a connection between a type A node and a type B node is determined by their distance in this space, rescaled by the product of their expected degrees. An additional parameter, $\beta$, acts as the inverse temperature of the ensemble: large $\beta$ values yield networks that are closely aligned with the underlying metric space, whereas smaller values increase randomness and recover the SCM in the limit $\beta \to 0$. Consequently, $\beta$ strongly influences clustering, which exhibits a topological phase transition at $\beta=1$~\footnote{The transition takes place at $\beta=D$ when working in a $D$-dimensional similarity space.}. Below this threshold, clustering vanishes in the thermodynamic limit; above it, it converges to a constant, reaching its maximum as $\beta \to \infty$. Both models are described in detail in {\it Materials and Methods}.

The mixture ensemble is defined as follows: Type B nodes are split into two groups, $B^{\mbox{\tiny{CM}}}$ and $B^{\mathbb{S}^1}$, with sizes $N_{\mbox{\tiny{B}}}^{\mbox{\tiny{CM}}}$ and $N_{\mbox{\tiny{B}}}^{\mathbb{S}^1}$, respectively. Group $B^{\mbox{\tiny{CM}}}$ is assigned expected degrees drawn from $\rho_{\mbox{\tiny{B}}}^{\mbox{\tiny{CM}}}(\kappa_{\mbox{\tiny{B}}})$, and its nodes connect to Type A nodes using the SCM connection probability given by~\eqref{eq:connection_probability_cm}. Similarly, group $B^{\mathbb{S}^1}$ is assigned expected degrees drawn from $\rho_{\mbox{\tiny{B}}}^{\mathbb{S}^1}(\kappa_{\mbox{\tiny{B}}})$, and its nodes connect to Type A nodes according to the bipartite-$\mathbb{S}^1$ model given by~\eqref{eq:connection_probability_s1}. Type A nodes are given an expected degree sequence drawn from $\rho_{\mbox{\tiny{A}}}(\kappa_{\mbox{\tiny{A}}})$. In our simulations, we used power-law distributions for the expected degrees of the form $\rho(\kappa)=(\gamma-1)\kappa_0^{\gamma-1}\kappa^{-\gamma}$ for $\kappa \ge \kappa_0$, with mean degree $\langle k \rangle=(\gamma-1)\kappa_0/(\gamma-2)$.

Figure~\ref{fig:results_synthetic} summarizes noise-detection performance on synthetic bipartite networks across parameter settings and significance levels $p$. In all experiments we fix $N_{\mbox{\tiny{A}}}=1000$, $\avg{k_{\mbox{\tiny{A}}}}=16$, and $\gamma_{\mbox{\tiny{A}}}=5$, yielding a comparatively homogeneous degree distribution for Type~A nodes, as often observed in empirical bipartite systems. For Type~B we set $N_{\mbox{\tiny{B}}}^{\mbox{\tiny{CM}}}=500$ and $N_{\mbox{\tiny{B}}}^{\mathbb{S}^1}=500$, each with $\avg{k_{\mbox{\tiny{B}}}}=8$. The values of $\gamma_{\mbox{\tiny{B}}}^{\mbox{\tiny{CM}}}$, $\gamma_{\mbox{\tiny{B}}}^{\mathbb{S}^1}$, and the geometric parameter $\beta$ used for Type~B are indicated in the figure.

The bars in Fig.~\ref{fig:results_synthetic} report the number of Type~B nodes labeled as noise. Light bars correspond to true positives (noisy nodes from $B^{\mbox{\tiny{CM}}}$ correctly identified), whereas dark bars correspond to false positives (signal nodes from $B^{\mathbb{S}^1}$ misclassified as noise), shown for different significance levels. Increasing $p$ generally increases sensitivity of the filter, leading to higher true-positive counts; this effect is especially pronounced for $p\ge 0.85$, where the method correctly retrieves more than $80\%$ of noisy nodes even under challenging conditions (low $\beta$ and highly heterogeneous Type~B degree distributions). Pushing $p$ closer to $1$ yields detection of nearly all noisy nodes but at the cost of a higher false-positive rate.

To quantify this trade-off, we report the F-Score, a standard metric for binary classification that balances precision and recall~\cite{vanRijsbergen1979,Manning2008IR}. It is defined in the range ($0,1$), with values close to 1 indicating better balance between identifying noise while avoiding false positives. The green squares in Fig.~\ref{fig:results_synthetic} show the F-Score for each configuration. The filter performs robustly, with $F\gtrsim 0.9$ for $p\ge 0.85$ at $\beta=1.5$. For $\beta=3$, the F-Score approaches $1$ as $p\to 1$. Additional sweeps over ensemble parameters and over different mixtures of $B^{\mbox{\tiny{CM}}}$ and $B^{\mathbb{S}^1}$ yield consistent trends; see the Supplementary Information for details.

\begin{figure}[t]
    \centering
    \includegraphics[width=\linewidth]{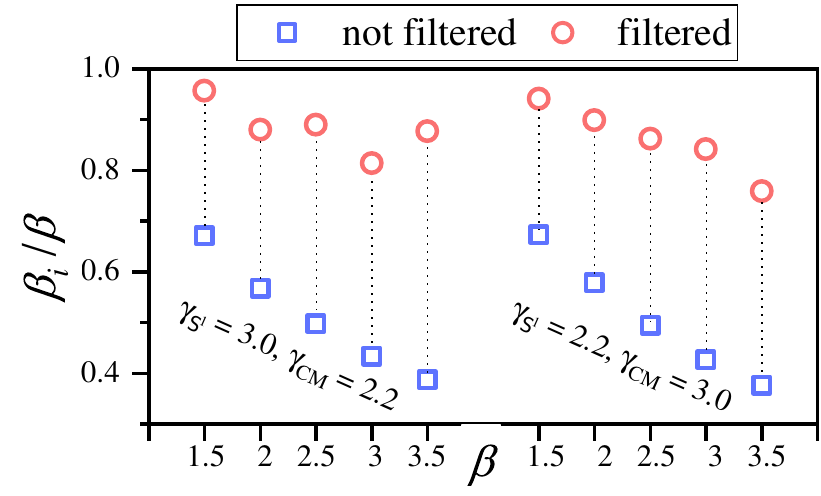} 
    \caption{Ratio between the inferred $\beta_i$ and the actual $\beta$  for different network configurations. Blue squares represent the unfiltered network, while pink circles correspond to the filtered network (significance level $p= 0.05$, or percentile $95\%$). The dotted lines highlight the improvement in $\beta_i$ after filtering. After filtering, the inferred $\beta_i$ approaches the true $\beta$, demonstrating the effectiveness of the filtering process. 
    }
    \label{fig:beta_ratio}
\end{figure}

As a final analysis of these synthetic experiments, we compare the parameter $\beta$ --as estimated by the B-Mercator embedding tool~\cite{Jankowski:2025er}-- for the original network, treating it as a realization of the bipartite-$\mathbb{S}^1$ model, with the corresponding estimate for the filtered network. In Fig.~\ref{fig:beta_ratio}, we observe that when the entire structure --geometric and noisy nodes together-- is considered, the inferred parameter $\beta_i$ is systematically lower than the ground-truth $\beta$ used to generate the connections of the nodes in the $B^{\mathbb{S}^1}$ set. After filtering, the inferred $\beta_i$ moves closer to the true value, despite the presence of some residual noise. This effect is summarized in Fig.~\ref{fig:beta_ratio}, where we report the ratio $\beta_i/\beta$ across different network parameters.

\subsection{Filtering of real networks}

\begin{figure}
\centering
{\includegraphics[width=\columnwidth]{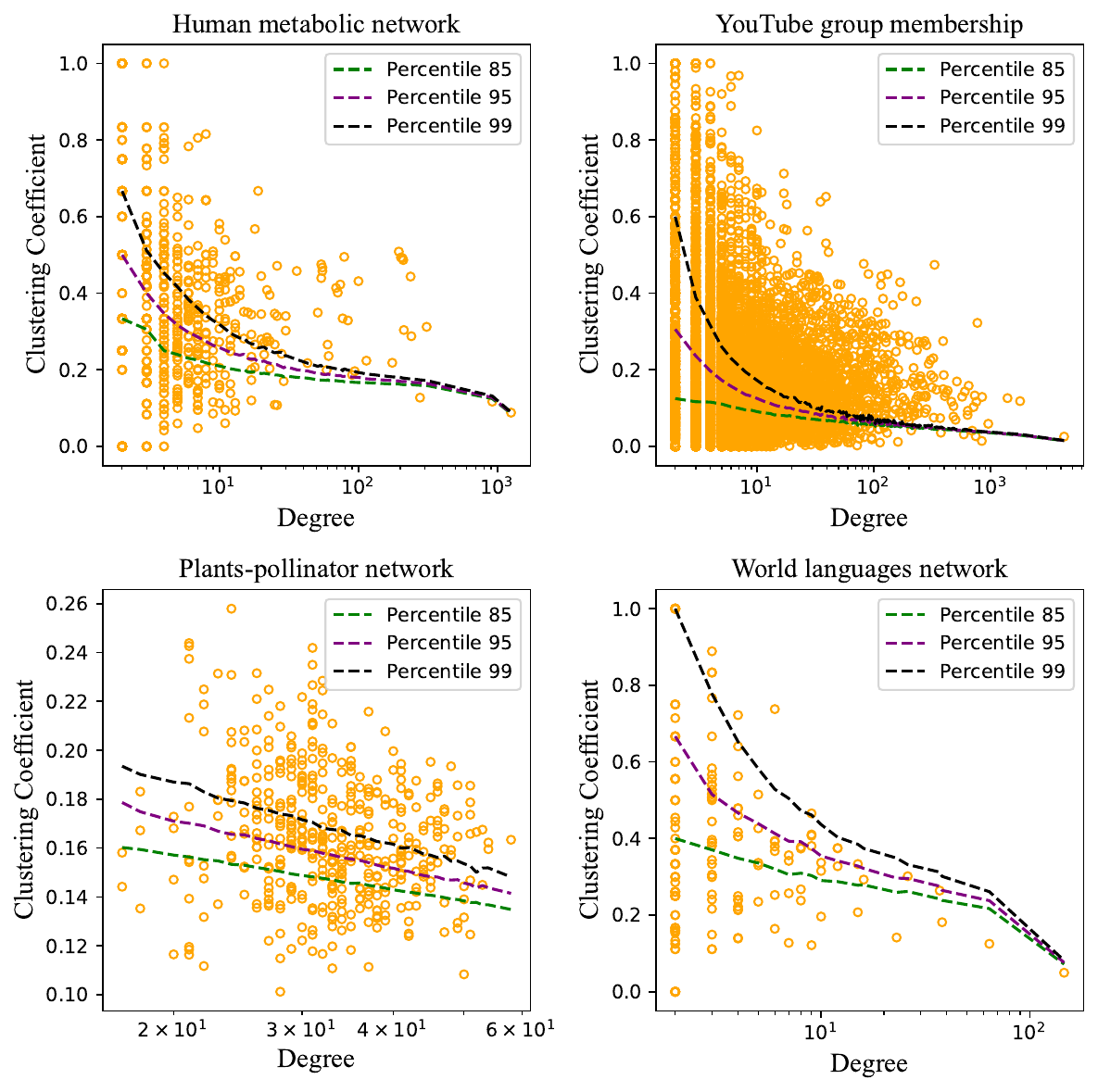}}
\caption{\textbf{Clustering coefficient versus degree across datasets.} Human metabolic network~\cite{BIGG,human_met}, YouTube membership network, plant-pollinator~\cite{PlantPollinator,PlantPollinatorII}, and world-language networks~\cite{languages}. Orange circles show clustering for type-\(B\) nodes; dashed lines mark the 85th (green), 95th (purple), and 99th (black) percentiles from degree-preserving randomizations. Percentiles serve as thresholds following Section~\ref{sec:filter}: points above are classified as geometrical (signal), and points below as non-geometrical (noise).}
\label{fig:results_real_networks}
\end{figure}

We analyze four real bipartite systems: (i) the human metabolic network from the BiGG database~\cite{BIGG}, linking reactions and metabolites in human cells; (ii) a YouTube group-membership network, linking users and groups; (iii) a plant-pollinator network~\cite{PlantPollinator}, linking plants to visiting insect species; and (iv) a world-language network, linking regions to their spoken languages~\cite{languages}. Table~\ref{table:REALNETWORKS} summarizes basic properties after removing degree-one nodes (clustering coefficients are undefined for such nodes); the reported sizes refer to the post-processing networks.

Figure~\ref{fig:results_real_networks} plots, for each dataset, the bipartite clustering coefficient of all type-\(B\) nodes versus degree, together with the 85th, 95th, and 99th percentiles from degree-preserving randomizations used as filtering thresholds (Section~\ref{sec:filter}), where type-\(B\) nodes are, respectively, metabolites, groups, plants, and languages. Across systems --metabolism, online group membership, plant-pollinator interactions, and world languages-- the same pattern emerges: many low-degree type-\(B\) nodes fall below the thresholds and are removed, consistent with near-random attachment, whereas nodes that participate in recurrent neighborhoods rise above the thresholds and are retained. Table~\ref{table:REALNETWORKS} reports the fraction of surviving type-\(B\) nodes at each \(p\)-value and shows the expected monotonic decrease with stricter thresholds.

The qualitative behavior is consistent across domains yet reveals domain-specific structure. In biochemical data, some ubiquitous components with very high degree (e.g., \(\mathrm{H_2O}\), coenzyme~A) are filtered out because their broad participation lacks recurrent co-occurrence, while other high-degree metabolites (e.g., ATP/ADP, NAD/NADP) remain because they concentrate within tightly reused reaction neighborhoods. In social affiliation, many small groups behave randomly and are removed, whereas large, structured communities persist due to strong co-membership clustering. In ecological interactions, a sizable share of plants are classified as randomly visited by insect species --suggesting broad pollinator mixing-- while repeatedly co-visited plants are retained. In linguistic data, widely dispersed languages are often filtered because cross-territorial ubiquity does not imply local co-use, but languages embedded in multilingual territories survive due to elevated local clustering. 

Together, these results show that the filter does more than suppress noise: it isolates geometrical signal --recurrent, locality-like structure-- while flagging non-geometrical components whose connections resemble random mixing. This integrated view holds across disparate bipartite systems, supporting the method's utility as a general tool for separating structure from noise.

\subsection{Structural properties of bipartite networks}

The two connectivity modes have clear implications for global structure. To assess structural robustness, we performed a percolation analysis that tracks the size of the giant connected component as we remove type-\(B\) nodes classified as geometric or as noise. After applying the filtering procedure, we partitioned type-\(B\) nodes into two sets: the \emph{geometric} set (nodes retained by the filter) and the \emph{noise} set (nodes removed by the filter). We then evaluated the resilience of the original bipartite network under targeted removals from each set and under random removals.

\begin{figure}[t]
\centering    
{\includegraphics[width=\columnwidth]{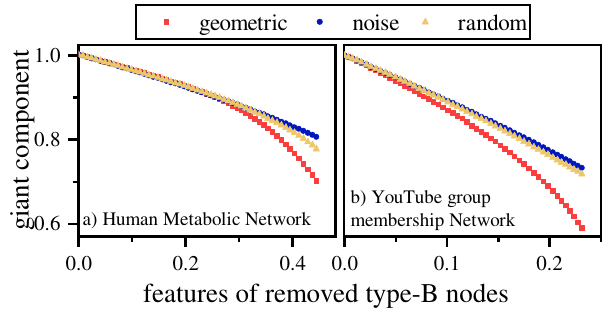}}
\caption{{\bf Percolation experiments.} Evolution of the giant connected component as type-\(B\) nodes are removed in (a) the human metabolic network and (b) the YouTube group-membership network. Each curve corresponds to removals from the geometric, noise, or random sets. For comparability, the number of removed nodes equals the size of the geometric set at \(p=0.05\). Results are averaged over 1{,}000 realizations.}\label{percolationHM} 
\end{figure}

As expected, removing geometric nodes produces a larger decline in the size of the giant component: once these structurally organized features are deleted, the remaining network is less coherently clustered. By contrast, removing the same number of noisy nodes has a weaker effect: the surviving network --now relatively more geometric-- better preserves large-scale connectivity. Figure~\ref{percolationHM} illustrates this behavior for the human metabolic and YouTube group-membership networks. Each curve reports the effect of removing nodes from the geometric, noise, or random sets. For comparability, the number of removed nodes in all cases equals the size of the geometric set at \(p=0.05\).

\begin{table*}[t] 
\caption{Summary of the real networks analyzed, including the number of nodes in Set A and Set B, and the sizes of the filtered subsets for different thresholds (p85, p95, and p99).
}\label{tab1}
\label{table:REALNETWORKS}
\begin{tabular}{@{}lllllllll@{}}
\toprule
\textbf{Network} & \textbf{Set A} & \textbf{$\avg{k}_{A}$}& \textbf{Set B} & \textbf{$\avg{k}_{B}$}& \textbf{C}  & \textbf{Filtered p85} & \textbf{Filtered p95} & \textbf{Filtered p99}\\
\midrule
Human Metabolic & $2212$ & $4.8$ & $1214$  & $8.6$ & $0.3932$ & $508(41.5\%)$ & $648(52.77\%)$ & $849(69.36\%)$\\
YouTube group membership & $37981$ & $3.9$ & $12009$ & $12.3$& $0.1015$  &$5366(44.68\%)$ & $7126(59.34\%)$ & $8900(74.11\%)$  \\
Plant Pollinator  & $1044$ & $14.6$ & $456$ & $33.5$ & $0.1312$ & $109(23.9\%)$ & $171(37.5\%)$ & $264(57.89\%)$ \\
World Languages   & $246$ & $3.8$  & $166$ & $5.6$& $0.1328$  & $79(47.59\%)$ & $123(74.1\%)$ & $159(95.78\%)$ \\
CORA & $2708$ & $18.18$  & $1431$ & $34.41$ & $0.1191$  & $587 (38.25\%)$ & $754(52.73\%)$ & $948(66.29\%)$ \\
Citiseer & $3312$ & $31.75$  & $3702$ & $28.4$ & 0.1144 & $608 (16.42\%)$ & $868 (23.44\%)$ & $1121 (30.27\%)$\\
BlogCatalog & $5196$ & $71.1$  & $8181$ & $45.11$ & $0.2606$  & $2877(35.16\%)$ & $3969(48.51\%)$ & $3969(48.51\%)$ \\

\hline 
\end{tabular}
\end{table*}

\section{Application to Neural Networks: the Impact of Filtration on Node Classification Tasks}

Modern neural networks are increasingly fed with graph-structured datasets in which vertices carry feature vectors and labels; improving the signal in these features can directly affect downstream classification.

Given a graph-structured dataset with \(N_n\) nodes and \(N_f\) features associated with the same set of nodes, we construct a bipartite network between nodes and features by treating features as entities, following~\cite{Roya}. Each node \(i\) has a binary feature vector \(\vec{f}_i \in \{0,1\}^{N_f}\), where each entry indicates the presence (\(1\)) or absence (\(0\)) of a feature. A node is therefore linked to all features for which \(\vec{f}_i=1\). We denote nodes by set \(A\) and features by set \(B\). Our goal is to apply the proposed filtration to set \(B\) to remove noisy features. As a baseline, we also create a \emph{randomly filtered} version of the data by removing at random the same number of features as in the filtration. We then feed the original, filtered, and randomly filtered feature matrices to two feature-based neural architectures --a multilayer perceptron (MLP) and hyperbolic neural networks (HNN)~\cite{HNN-Ganea}-- to perform node classification.

\begin{figure}[t]
\centering    
{\includegraphics[width=\columnwidth]{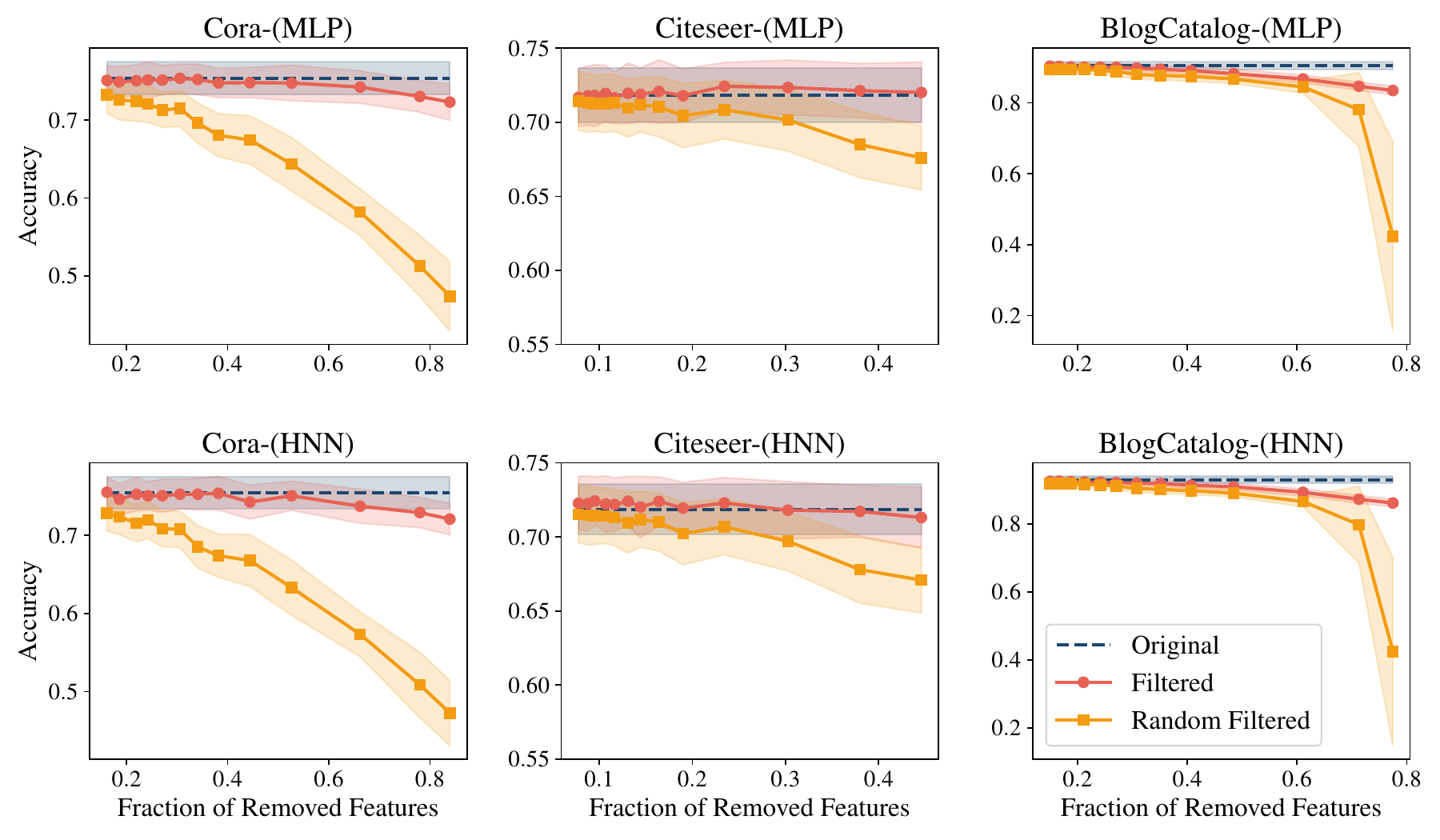}}
\caption{Node-classification accuracy as a function of the fraction of removed features for two feature-based neural networks (MLP and HNN). Models are trained on three versions of the node-feature bipartite data: the original features, a filtered set obtained with our node-filtration method, and a randomly filtered set in which the same number of features is removed at random. Error bars show \(\pm1\sigma\) around the mean over 100 random train/validation/test splits; for randomly filtered data they also incorporate variability across 10 independent random filtrations. In all experiments, nodes are split into training (70\%), validation (15\%), and test (15\%) sets. Models are optimized with Adam~\cite{adam2014}, use two layers with hidden dimension 16, and are trained with dropout \(=0.2\) and weight decay \(=0.001\). Class labels are assigned via a softmax over the final layer.}
\label{fig:NN-Results} 
\end{figure}

Figure~\ref{fig:NN-Results} reports node-classification accuracy (fraction of correctly classified nodes) on three real datasets --the Cora~\cite{CORA}, Citeseer~\cite{CITISEER}, and BlogCatalog~\cite{BlogCatalog} datasets  -- for the two neural architectures. In all cases, as the filtration threshold \(p\) increases and features identified as noise are removed, accuracy remains nearly unchanged and close to that of the unfiltered data. In contrast, randomly filtering features produces a pronounced drop in accuracy, especially under large random removals. The Cora dataset is illustrative: node labels are highly correlated with geometric features and removing \(80\%\) of the random features leaves accuracy essentially unchanged relative to the original network, whereas removing the same number of features chosen at random reduces accuracy to below \(50\%\).

These results indicate that our filtration effectively extracts the structural backbone of the node-feature bipartite network while discarding features that do not contribute to predictive performance. Taken together, these results suggest that one can retain a small, relevant subset of features without sacrificing accuracy, thereby substantially reducing the computational cost of neural-network training.

\section{Discussion}

We introduced a statistically grounded filter that disentangles geometrical (structured) from random connectivity in bipartite networks by benchmarking node-level clustering against degree-preserving randomizations. Across synthetic and empirical systems, the same picture emerges: a relatively small geometric backbone concentrates recurrent, locality-like neighborhoods, while a sizable random-like component accounts for diffuse mixing. Rather than treating this latter part as nuisance, our results show that it carries domain-specific meaning (e.g., broad pollinator mixing) and should be interpreted in context.

Validation on controlled mixtures (SCM + bipartite-$\mathbb{S}^1$ model) shows high detection quality for practical thresholds (F-scores typically $\gtrsim 0.9$ for percentiles above $85\%$) and a marked improvement in geometric parameter inference after filtering, indicating that the method sharpens latent-space structure without assuming a specific generative model for the signal. On real networks --including metabolism, online affiliation, plant-pollinator interactions, and languages-- the filter consistently retains entities embedded in recurrent neighborhoods, while broadly ubiquitous or weakly co-occurring entities fall below the randomization threshold. Together, these findings indicate that the approach extracts a compact geometric backbone that better captures the organizing principles of the data, while leaving a residual random-like layer whose role depends on the domain (stabilizing diversity vs.\ adding noise).

The separation has concrete structural and functional implications. Percolation experiments reveal that targeted removal of geometric nodes disproportionately fragments connectivity relative to removing the same number of noisy nodes, consistent with the backbone's role in holding the system together. In node-feature graphs for machine learning, discarding features flagged as noise preserves classification accuracy, whereas removing an equal number of features at random degrades performance. Thus, the filter reduces dimensionality and computational cost while keeping predictive content intact, positioning it as a complementary pre-processing tool for graph-based pipelines.

Several limitations suggest directions for follow-up work. First, our discrimination relies on clustering conditioned on degree and a the CM as a null model for noise; networks with strong constraints beyond degree may require adapted null models. Second, clustering is a sufficient but not exclusive signal of geometry; integrating additional local or mesoscopic statistics could refine the boundary between signal and noise. Finally, many systems are weighted, directed, temporal, or multilayer; extending the filter to these settings --and to both node sets simultaneously-- should broaden applicability.

In sum, the method offers a simple, scalable route to parse mixed mechanisms in bipartite graphs. By isolating a geometric backbone and characterizing the complementary random-like layer, it clarifies which components drive organization and which provide flexible mixing. This separation improves latent-space inference, strengthens robustness analyses, and streamlines downstream learning, providing a practical bridge between descriptive network analysis and actionable modeling across ecological, biological, technological, and social systems.

\section*{Materials and Methods}
\subsection*{Evaluation metric}
We quantify detection performance with the F-score, the harmonic mean of precision and recall. Let $TP$, $FP$, and $FN$ denote true positives, false positives, and false negatives, respectively:
\begin{equation}
\label{eq:FScore}
F=\frac{2TP}{2TP+FP+FN},
\end{equation}
with $0\le F\le 1$.

\subsection*{Synthetic bipartite benchmarks}
We generate bipartite graphs with node sets $A$ and $B$ in which only a subset of $B$ is contaminated by noise. Set $B$ is a mixture of geometric nodes $B^{\mathbb{S}^1}$ and random nodes $B^{\mathrm{CM}}$, so that $N_B=N_B^{\mathbb{S}^1}+N_B^{\mathrm{CM}}$. Nodes in $A$ and in $B^{\mathbb{S}^1}$ carry hidden variables $(\kappa,\theta)$, where $\kappa$ controls expected degree and $\theta\in[0,2\pi)$ is an angular similarity coordinate.

\paragraph*{$\mathbb{S}^1$ (geometric) links.}
Edges between $A$ and $B^{\mathbb{S}^1}$ are drawn independently with probability
\begin{equation}
\label{eq:connection_probability_s1}
p_{\mathbb{S}^1}(\kappa_A,\kappa_B,\Delta\theta)=\left[1+\left(\frac{R\,\Delta\theta}{\mu_{\mathbb{S}^1}\kappa_A\kappa_B}\right)^{\beta}\right]^{-1},
\end{equation}
where $\Delta\theta$ is the angular distance, $\beta>1$ controls geometric sharpness, and $\mu_{\mathbb{S}^1}$ sets the mean degrees. Details of both models are in the Supplementary Information.

\paragraph*{Configuration-model (random) links.}
Edges between $A$ and $B^{\mathrm{CM}}$ follow a soft configuration model,
\begin{equation}
\label{eq:connection_probability_cm}
p_{\mathrm{CM}}(\kappa_A,\kappa_B)=\frac{\mu_{\mathrm{CM}}\kappa_A\kappa_B}{1+\mu_{\mathrm{CM}}\kappa_A\kappa_B},
\end{equation}
with $\mu_{\mathrm{CM}}$ fixed to match the desired mean degrees in $A$ and $B^{\mathrm{CM}}$.

\paragraph*{Degree distributions.}
Unless otherwise noted, hidden degrees are sampled from power laws
$\rho(\kappa)\propto \kappa^{-\gamma}$ with lower cutoffs chosen so that the target mean degrees exist. Exponents for $A$, $B^{\mathbb{S}^1}$, and $B^{\mathrm{CM}}$ may differ.

\subsection*{Bipartite clustering coefficient}
We measure local geometric organization with a bipartite clustering coefficient for node $i$,
\begin{equation}
\label{eq:clustering}
C_3(i)=\frac{2\,\tilde T_i}{k_i(k_i-1)},\qquad
\tilde T_i=\sum_{\alpha\neq\beta}\frac{T_{\alpha\beta}(i)}{\min(k_\alpha,k_\beta)-1}.
\end{equation}
Here $k_i$ is the degree of $i$; $T_{\alpha\beta}(i)$ counts the number of common neighbors of the neighbor pair $(\alpha,\beta)$ in the opposite set; and the denominator normalizes by the maximum number of possible common neighbors for that pair. This definition weights pairs more heavily when they are closed through many shared neighbors.

\section*{Acknowledgements}{We acknowledge support from: Grant TED2021-129791B-I00 funded by MCIN/AEI/10.13039/501100011033 and the ``European Union NextGenerationEU/PRTR''; Grant PID2022-137505NB-C22 funded by MCIN/AEI/10.13039/501100011033; M. B. acknowledges the ICREA Academia award, funded by the Generalitat de Catalunya}

\section{References}


%

\end{document}